\documentclass[12pt,aps,superscriptaddress,prl]{revtex4}
\usepackage{amsmath}
\usepackage{amsthm}
\usepackage[colorlinks=true,urlcolor=blue,citecolor=blue,linkcolor=blue]{hyperref}
\usepackage[shortlabels]{enumitem}
\usepackage{xcolor, graphicx, ulem}
\usepackage{amsfonts}
\usepackage{graphicx}
\usepackage{bbold}
\usepackage{stmaryrd}
\usepackage{color}
\usepackage{hyperref}
\usepackage{verbatim}
\usepackage{cases}
\usepackage{amsfonts}
\usepackage{amssymb}
\usepackage[]{mathrsfs}
\usepackage{xcolor}
\usepackage{anysize}
\usepackage{epsfig}
\usepackage{bm}
\usepackage{setspace}
\usepackage{enumerate}
\usepackage{dcolumn}
\usepackage{bm}
\usepackage{enumitem}
\usepackage{soul}

\newcommand{\bra}[1]{\ensuremath{\left\langle#1\right|}}
\newcommand{\ket}[1]{\ensuremath{\left|#1\right\rangle}}
\expandafter\let\csname equation*\endcsname\relax
\expandafter\let\csname endequation*\endcsname\relax

\usepackage{color}
\usepackage{mathtools}
\usepackage{pgfplots}
\usepackage{braket}
\usepackage{stackengine}
\usepackage{geometry}
\newgeometry{vmargin={15mm}, hmargin={25mm,25mm}}

\begin{document}

\title{Two-photon edge states in photonic topological insulators:\\ topological protection versus degree of entanglement}

\date{\today}

\author{Konrad Tschernig}\email{konrad.tschernig@physik.hu-berlin.de}
\affiliation{Max-Born-Institut, Max-Born-Stra{\ss}e 2A, 12489 Berlin, Germany}
\affiliation{Humboldt-Universit\"at zu Berlin, Institut f\"{u}r Physik, AG Theoretische Optik \& Photonik, Newtonstra{\ss}e 15, 12489 Berlin, Germany}

\author{\'{A}lvaro Jimenez-Gal\'{a}n}
\affiliation{Max-Born-Institut, Max-Born-Stra{\ss}e 2A, 12489 Berlin, Germany}

\author{Demetrios~N.~Christodoulides}
\affiliation{CREOL, The College of Optics and Photonics, University of Central Florida, , Orlando, FL 32816-2700, USA}

\author{Misha~Ivanov}
\affiliation{Max-Born-Institut, Max-Born-Stra{\ss}e 2A, 12489 Berlin, Germany}
\affiliation{Humboldt-Universit\"at zu Berlin, Institut f\"{u}r Physik, AG Theoretische Optik \& Photonik, Newtonstra{\ss}e 15, 12489 Berlin, Germany}
\affiliation{Blackett Laboratory, Imperial College London, London, UK}

\author{Kurt Busch}
\affiliation{Max-Born-Institut, Max-Born-Stra{\ss}e 2A, 12489 Berlin, Germany}
\affiliation{Humboldt-Universit\"at zu Berlin, Institut f\"{u}r Physik, AG Theoretische Optik \& Photonik, Newtonstra{\ss}e 15, 12489 Berlin, Germany}

\author{Miguel~A.~Bandres}\email{bandres@creol.ucf.edu}
\affiliation{CREOL, The College of Optics and Photonics, University of Central Florida, , Orlando, FL 32816-2700, USA}

\author{Armando Perez-Leija}\email{apleija@gmail.com}
\affiliation{Max-Born-Institut, Max-Born-Stra{\ss}e 2A, 12489 Berlin, Germany}
\affiliation{Humboldt-Universit\"at zu Berlin, Institut f\"{u}r Physik, AG Theoretische Optik \& Photonik, Newtonstra{\ss}e 15, 12489 Berlin, Germany}

\begin{abstract}
\textbf{Topological insulators combine insulating properties in the bulk with scattering-free transport along edges, supporting dissipationless unidirectional energy and information flow even in the presence of defects and disorder. 
The feasibility of engineering quantum Hamiltonians with photonic tools, combined with the availability of entangled photons, raises the intriguing possibility of employing topologically protected entangled states  in optical quantum computing and information processing.  However, while two-photon states built as a product of two 
topologically protected single-photon states inherit full protection from their single-photon “parents”, high degree of non-separability may 
lead to rapid deterioration of the two-photon states after propagation through disorder. We identify physical mechanisms which contribute to the vulnerability 
of entangled states in topological photonic lattices and
present clear guidelines for maximizing entanglement without sacrificing  
topological protection.}
\end{abstract}
\maketitle
\newpage


The prospect of generating topologically protected entangled states of several photons is a highly intriguing proposition \cite{Rechtsman:16,Mittal:16,Gneiting2019}.  Specifically, topological protection can enable robust transport of quantum information across disordered photonic structures without degradation \cite{Blanco-Redondo568,Wang:19}, just as efficiently as for single-particle wavepackets \cite{Hafezi2013,BlancoRedondo2016,Cheng2016,Khanikaev2017,Jotzu2014}. \\
In recent years, we have witnessed several experimental demonstrations of topological protection at the single-photon level in integrated one-dimensional lattice systems.
Notably, Wang and co-workers showed that the fundamental quantum features of spatially entangled biphoton-states can be protected against disorder in the so-called Su-Schrieffer-Heeger (SSH) topological lattice \cite{Topologicallyprotectedentangledphotonicstates}. 
Interestingly, SSH lattices turned out to be equally effective in protecting polarization-entangled photon pairs  \cite{wang2019topologically}. 
Another important ingredient was provided by Tambasco \textit{et~al} in Ref. \cite{tambasco2018quantum} showing that Hong-Ou-Mandel two-photon interference of topological edge-modes is feasible, by implementing a topological beamsplitter in a judiciously engineered time-dependent Harper-model.\\
Concurrently, on the theory front several ideas have been suggested to investigate topological two-photon effects in linear \cite{Gorlach2017,Gorlach2018} and nonlinear \cite{Gorlach2017-2} lattice systems. In this regard, an intriguing proposition was recently put forward in Ref. \cite{PhysRevA.102.013510}, where the Bose-Hubbard model, which is topologically trivial for single particles, becomes topologically nontrivial for two interacting photons. 
That is, particle interactions have a dramatic impact on topological properties, not only modifying the topology of the spectra but also creating a topological order in otherwise topologically trivial systems.\\
In order to maximize the potential of topological photonic networks for transferring quantum information, it is indispensable to have a considerable number of edge modes at our disposal. One possibility is to use two-dimensional topological systems, which intrinsically support a multitude of topological edge-states \cite{Barik666,Bandreseaar4005,Hararieaar4003}.\\

In two-dimmensional photonic topologial insulators, single particle edge-states reside in the gap existing between the energy bands supporting the bulk states \cite{Kane05,Haldane2008,Feng2017}. 
Thus, breaking the topological protection requires disorder with sufficient strength to close the bandgap.
For states describing two indistinguishable photons, the same bandgap is fundamentally lacking. 
The reason is because the propagation eigenvalues $\lambda^{(2)}_{12}$
for two-photon eigenstates in a photonic system are given by the sum of the eigenvalues $\lambda_{1}, 
\lambda_{2}$ corresponding to the constituent individual photons, 
$\lambda^{(2)}_{12}=\lambda_{1}+\lambda_{2}$. 
This implies that we can keep $\lambda^{(2)}_{12}$ constant while increasing
$\lambda_{1}$ and simultaneously decreasing $\lambda_{2}$,
or vice versa.
In this way, we can combine two single-photon bulk states, one from the lower and one from the upper band, to create a biphoton bulk-bulk state whose energy lies inside the single particle bandgap. 
This fundamental additive property of the single particle eigenvalues removes the bandgap and leads to massive degeneracies of the edge-edge, edge-bulk, and bulk-bulk two-photon states.
Hence, considering the lack of the topological bandgap for two-photon systems, it is not clear whether topological protection will be automatically granted to two-particle states provided the constituent single particles are topologically protected.\\

In solids, the degeneracies described above
lead to the decay of two-electron edge states when electron-electron correlation are substantial \cite{Strunz2020,Stuhler2020,Anirban2019}. This
decay mechanism is reminiscent of auto-ionization, where electron-electron
correlation leads to energy exchange between the two particles, coupling 
two bound electronic excitations to an energy-degenerate 
bound-continuum two-electron state \cite{Fano1961,Galan2014}.
Still, photonic systems are fundamentally different from solids,
as the two photons do not readily interact with each other \cite{Saleh2000}.
Consequently, the evolution operator for two-photon states, $U^{(2)}(z)$, breaks down into the product of
 two propagators for individual single-photon states, $U^{(2)}(z)=U(z)\otimes U(z)$ \cite{Nielsen2010}. 
Thus, a natural question to ask is whether such a factorization and the absence of bangap will prevent decoherence and dissipation of non-factorizable two-photon edge-states into the bulk?
To address this question we analyze possible mechanisms of dissipation of two-photon edge-edge states into the bulk of the system.\\

In lattice systems, static disorder can be introduced in either the site energies  -- termed diagonal disorder \cite{Lahini2008}-- or in the coupling coefficients --so-called off-diagonal disorder \cite{Martin2011}. In either case, static disorder is represented by a single-particle 
operator $\hat V^{(1)}$.  Since such perturbation 
is time-independent,  energy conserving 
{\it resonant} coupling into the bulk is absent within 
first-order perturbation theory -- the single-particle 
transition induced by $\hat V^{(1)}$ does not preserve energy.
The process that can resonantly couple a two-photon edge-edge 
state to a bulk-bulk, or to a bulk-edge, state would require a correlated change of 
states for both photons and  it might arise within the 
second order corrections in $\hat V^{(1)}$. \\
To see this we examine the second-order transition matrix elements 
between an initial two-photon edge-edge state 
$|i\rangle=|n_i,m_i\rangle$ and a final edge-bulk, or bulk-bulk, 
state $|f\rangle=|n_f,m_f\rangle$
\begin{equation}
V^{(2)}_{fi}=\sum_{j'} \frac{V^{(1)}_{fj'}V^{(1)}_{j'i}}
{\lambda^{(2)}_{j'}-\lambda^{(2)}_i},
\label{eq:SecondOrderMatrixElement}
\end{equation}
where $|j'\rangle=|n',m'\rangle$ labels intermediate {\it virtual states} and
$\lambda^{(2)}_{j'}=\lambda_{n'}+\lambda_{m'}$.
The single-particle nature of $\hat V^{(1)}$ ensures that only two
terms corresponding to the two possible time-orderings 
of the two single-particle transitions are left in the sum
\begin{equation}
V^{(2)}_{fi}=V^{(1)}_{n_f,n_i}V^{(1)}_{m_f,m_i}
\left[
\frac{1}{\lambda_{n_f}-\lambda_{n_i}}
+
\frac{1}{\lambda_{m_f}-\lambda_{m_i}}
\right].
\label{eq:SecondOrderMatrixElement2}
\end{equation}
The stationary nature of the disorder dictates that real 
transitions from the edge-edge states can occur only if the initial eigenvalue $\lambda^{(2)}_i=\lambda_{n_i}+\lambda_{m_i}$ 
is equal to the final one $\lambda^{(2)}_f=\lambda_{n_f}+\lambda_{m_f}$.
Therefore, $\lambda_{n_f}-\lambda_{n_i}=-(\lambda_{m_f}-\lambda_{m_i})$, 
and the two terms in $V^{(2)}_{fi}$  exactly cancel each other, $V^{(2)}_{fi}=0$. That is, two-particle dissipation from  each product
state is zero, and the same is true for any entangled  two-particle state $\psi^{(2)}$  represented as a 
quantum superposition of two possible distinguishable configurations $\psi_{i}^{(2)}$, e.g. $\psi^{(2)}=\psi_{1}^{(2)}+\psi_{2}^{(2)}$.
Physically, this destructive interference is a direct outcome
of the non-interacting nature of indistinguishable photons, which ensures
that the two-particle eigenvalue is a sum of the two single-particle
eigenvalues, and that the two-particle propagator is a product
of the single-particle counterparts. 

In contrast to dissipation, the situation with dephasing can be different:
while each constituent state in an entangled superposition
can be protected against disorder \cite{Rechtsman:16}, the overall superposition is, in general, not.
To be precise, motion through different disordered regions may lead to 
disorder-induced random phase shifts between the states
destroying the entanglement. 
To avoid this fate, all states in the superposition must travel across the same spatial region
of the photonic structure, such that they are affected by disorder in the exact same manner \cite{LeijaNPJ}.
These effects have been explored for spatial path-entangled states \cite{Rechtsman:16}, and for states built from an entangled superposition of an initial non-stationary state $\psi_{1}^{(2)}$ with its time-delayed replica $\psi_{1}^{(2)}(\tau)$, that is, entangled states in the time-domain \cite{Mittal:16}. 
%
In these two cases, however, the entanglement of the states can be related to the entanglement of symmetrized wavefunction of identical particles \cite{TschernigEP}. Consequently, the states exhibit the lowest possible amount of entanglement, as indicated by the corresponding Schmidt numbers $S_{N}=2$. Throughout this work we use the Schmidt number to quantify the amount of entanglement: $S_{N}=1$ denotes complete separability while $S_{N}\gg1$ corresponds to high entanglement \cite{Sperling_2011}.\\
A more appealing type of highly entangled two-photon states are multimode optical Gaussian states in which both photons are most likely to be found inhabiting   any waveguide, within an excitation window, simultaneously \cite{Humphreys}. The importance of such states is based on the fact that any phase difference arising among the paths becomes enhanced by a factor of two in comparison with single photon states \cite{Afek879}. 
Naturally, the enhanced phase sensitivity of such highly entangled two-photon states manifests as faster diffraction of the associated wave-packets propagating in any photonic system, periodic and disordered \cite{Giuseppe2013}. Therefore, it is not clear to what extent topological protection will persist for these types of highly entangled states.

\begin{figure}[b!]
\begin{center}
\includegraphics[width=0.8\linewidth]{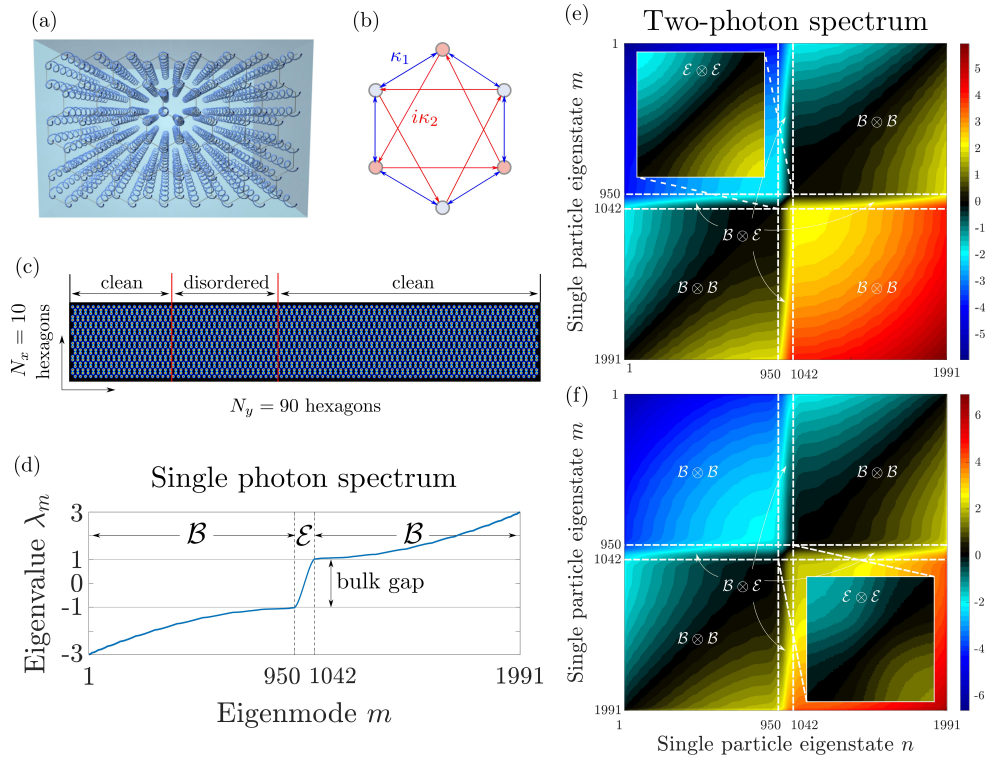}
\caption{(a) A photonic Haldane lattice system implemented using a honeycomb lattice of helical waveguides. 
(b) Elementary hexagonal cell of the Haldane system, with real-valued 
nearest-neighbour coupling (blue arrows) 
$\kappa_1=1$ and imaginary next-nearest-neighbour coupling (red arrows);
$\kappa_2=i\kappa_1/5$ along the arrow and $-i\kappa_1/5$ in the opposite direction. In (c) we depict the lattices used in our simulations.
Panel (d) shows single-photon spectrum formed by eigenvalues
$\lambda_n$. Panels (e) and (f) show the 
two-photon eigenspectra, (e) without and (f) with disorder. 
Colors encode the two-photon eigenvalue $\lambda^{(2)}_{n,m}=\lambda_n + \lambda_m$. For better appreciation we have separated the spectrum into the 
subspaces edge-edge $\left(\mathcal{E}\otimes\mathcal{E}\right)$, 
bulk-edge $\left(\mathcal{B}\otimes\mathcal{E}\right)$, 
and bulk-bulk $\left(\mathcal{B}\otimes\mathcal{B}\right)$.
}
\label{fig:hexagon}
\end{center}
\end{figure}

\begin{figure}[t!]
\begin{center}
\includegraphics[width=0.8\linewidth]{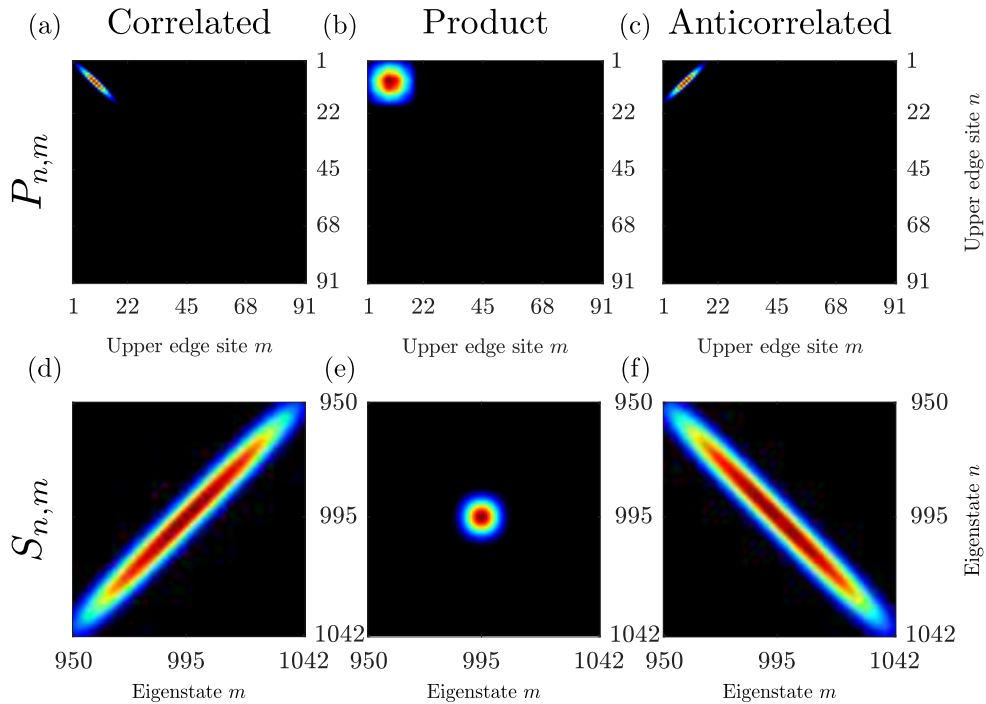}
\caption{Spatial (top row) and spectral 
(bottom row) correlation maps for different two-photon states. (a, d) Correspond to strongly 
spatially correlated states $\ket{\psi^{(2)}_c}$, $(\sigma_c,\sigma_a)=(\sqrt{40},0.01)$, 
(b, e) are for product states $\ket{\psi^{(2)}_p}$, $(\sigma_c,\sigma_a)=(\sqrt{40},\sqrt{40})$, and 
(c, f) for strongly anti-correlated state $\ket{\psi^{(2)}_a}$, $(\sigma_c,\sigma_a)=(0.01,\sqrt{40})$. 
}
\label{fig:twophotonstates}
\end{center}
\end{figure}

In what follows we analyze the impact of disorder onto a continuum of two-photon states that extend from the correlated to the anti-correlated limits, passing through a completely separable state. For our analysis we consider two topological lattices, one periodic and one aperiodic. In the periodic case we consider the Haldane model \cite{Haldane1988}, and for the aperiodic we use a square lattice whose single-particle dynamics corresponds to the quantum Hall effect \cite{Hafezi2013,HafeziNatPhys}. The results for the Haldane model are presented here, while the quantum Hall effect lattice is discussed in the supplementary section V.\\ 
In optics, a first order approximation of the Haldane model can be implemented using a honeycomb lattice  composed of helical waveguides as illustrated in Fig.~(\ref{fig:hexagon}-a), see pioneering work \cite{Rechtsman2013}. 
In this system, every waveguide has a nearest-neighbor coupling 
$\kappa_{1}$ and a complex second-nearest-neighbor coupling $\kappa_{2}$ or $\kappa_{2}^{*}$, see Fig.~(\ref{fig:hexagon}-b). 
At the single-photon level, the Haldane lattice is governed by the Hamiltonian \cite{Rechtsman:16}
\begin{align}
\hat{H} =  \sum_{i}\beta_{i}\hat{a}^\dagger_i \hat{a}_i + \kappa_1 \sum_{\langle i,j \rangle} \left( \hat{a}^\dagger_i \hat{a}_j + \hat{a}^\dagger_j \hat{a}_i\right) + i\kappa_2 \sum_{\langle \langle i,j \rangle \rangle} \left( \hat{a}^\dagger_i \hat{a}_j - \hat{a}^\dagger_j \hat{a}_i \right),
\label{eq:SPH}
\end{align}
where $\beta_{i}$ represents the propagation constant of the $i$-th waveguide and the corresponding optical mode is represented by the creation 
(annihilation) operator, $\hat{a}^\dagger_i$ ($\hat{a}_i$). Notice, in a disorder-free lattice $\beta_{i}=\beta$.
The symbols $\langle  \rangle$ and $\langle \langle \rangle \rangle$ indicate 
summation over nearest and next-nearest neighbor sites, respectively. 
The lattice used in our simulations is a ribbon with $N_y=90$ hexagons in the $y$-direction 
and $N_x=10$ hexagons in $x$-direction, Fig.~(\ref{fig:hexagon}-c). We normalized the units in terms of $\kappa_1$ throughout this work, and set $\kappa_2=i\kappa_1/5$.\\
For pure states of two indistinguishable noninteracting particles the Hamiltonian is $H_{2}=H \otimes I+I\otimes H$, where $H$ is the single-particle Hamiltonian and $I$ is the identity operator \cite{Leija:15}. The two-photon eigenstates are given by the symmetric tensor-product combinations of the single-photon eigenstates
\begin{equation}
\ket{\phi^{(2)}_{m,n}}=\left\{ \begin{matrix}
\ket{\phi_m} \otimes \ket{\phi_n} \Leftrightarrow  m=n, \\
\frac{1}{\sqrt{2}} \left( \ket{\phi_m} \otimes \ket{\phi_n} + \ket{\phi_n} \otimes \ket{\phi_m} \right)\Leftrightarrow  m \neq n .
\end{matrix} \right.
\label{eq:twophotoneigenstates}
\end{equation}
As alluded to above, the two-photon eigenvalues are the sums of the 
single-photon ones, $\lambda^{(2)}_{m,n} = \lambda_m + \lambda_n$. \\
In the absence of disorder, the eigenvalue spectrum for single-photon 
states in a finite lattice exhibits topological edge states in the bandgap \cite{Ozawa2019},  Fig.~(\ref{fig:hexagon}-d). 
In contrast, for two indistinguishable photons, the  spectrum does not have a bandgap:
the edge-edge states can have the same eigenvalues $\lambda^{(2)}_{n,m}=\lambda_n + \lambda_m$ as those lying in the bulk-bulk region,
Fig.~(\ref{fig:hexagon}-e). \\
To include disorder, we separate the lattice shown in Fig.~(\ref{fig:hexagon}-c) into three regions \cite{Rechtsman:16}. 
The left and right parts of the system are disorder-free, while its middle part 
exhibits diagonal disorder \cite{Lahini2008}, that is, random modifications of the on-site refractive index taken from a
normal distribution with zero mean and variance $\sigma=1$. 
Importantly, taking $\sigma=1$ ensures that the disorder strength  
does not destroy the topological protection for single photons, since $\sigma=1$ corresponds to half the size of the topological bandgap.
The two-photon eigenspectrum in the presence of disorder is shown 
in Fig.~(\ref{fig:hexagon}-f).\\
We now send  trial two-photon wavepackets into the system. 
They are built from single-photon edge states and vary continuously from 
an unentangled product state, with Schmidt number 
$S_{N}=1$, to highly-entangled two-photon states, 
$S_{N}\gg 1$ \cite{Ekert1995,Law2000}, with the two photons either correlated or anti-correlated in space \cite{CorrAnticorr}.\\ 
To construct these states, we begin with protected single-photon states as a template,
%
$\ket{\tilde{\psi}^{(1)}_{\sigma}}=\sum_{j=1}^{M_e}(-1)^{j} 
e^{-\frac{\left(x_0-j\right)^2}{2 \sigma^2}}\ket{j}$,
%
where $\ket{j}$ describes a photon initialized in waveguide $j$, 
$M_e=20$ is the selected range of waveguides
in the upper left edge of our system, and $x_0=(M_e+1)/2=10.5$ is the center of this
range. These single-photon wavepackets travel through both clean and disordered lattice
without scattering to the bulk or back scattering. Importantly, the alternating sign $(-1)^j$ in the 
amplitude ensures that the wavepacket has 
proper momentum and  resides in the single-photon edge subspace.\\ 
Next, we construct our trial two-photon states as follows
\begin{equation}
\ket{\tilde{\psi}^{(2)}_{\sigma_c,\sigma_a}}=\sum_{j,k=1}^{M_e} \psi_{j,k}\ket{j, k}=\sum_{j,k=1}^{M_e}(-1)^{j+k} e^{-\frac{(j-k)^2}{4 \sigma_a^2}-\frac{\left(x_0-(j+k)/2\right)^2}{\sigma_c^2}}\ket{j, k}.
\label{eq:EG}
\end{equation}
%
Here, $\ket{j, k}$ represents the state where a photon starts at waveguide $j$ and its twin at $k$.
The spatial two-photon correlations are controlled by the parameters $\sigma_c$ and $\sigma_a$.
For $\sigma_c\gg \sigma_a$ we have a spatially correlated state, in which both photons most probably enter into the same waveguide simultaneously \cite{Giuseppe2013}. 
For $\sigma_a\gg \sigma_c$ we obtain a spatially anti-correlated state, in which the two photons enter at two waveguides symmetrically lying on opposite sides of the window covered by the wavefunction \cite{CorrAnticorr}.\\
Finally, we must ensure that the initial wavepackets only include edge states.
To this end,  we project our state onto the two-photon eigenstates $\ket{\phi^{(2)}_{m,n}}$ 
of the system and then remove the components belonging to the subspaces 
$\mathcal{B}\otimes\mathcal{E}$ and $\mathcal{B}\otimes\mathcal{B}$, keeping only states that belong to the
edge-edge subspace
\begin{equation}
\ket{\psi^{(2)}_{\sigma_c,\sigma_a}}=\frac{1}{A}\sum_{m,n}^{\mathcal{E}\otimes\mathcal{E}} 
\sum_{j,k=1}^{M_e} \psi_{j,k} \braket{\phi^{(2)}_{m,n}|j,k} \ket{\phi^{(2)}_{m,n}},
\label{eq:7}
\end{equation}
where $A$ is the normalization constant. It is worth noting that two-photon states described by Eq.~\eqref{eq:7} are a lattice adoption of Gaussian two-mode squeezed states \cite{Schneeloch_2016}, which are a commonplace choice in quantum optical experiments.
The corresponding spatial
$P_{j,k}=|\braket{j, k|\psi^{(2)}_{\sigma_c,\sigma_a}}|^2$
and  spectral 
$S_{m,n}=|\braket{\phi^{(2)}_{m,n}|\psi^{(2)}_{\sigma_c,\sigma_a}}|^2 $
correlation maps  of our initial states, Eq. (\ref{eq:7}), are shown in Fig.~(\ref{fig:twophotonstates}).
Tuning $\sigma_a$ and $\sigma_c$, one can go from  the spatially correlated state 
$\ket{\psi^{(2)}_c}$, Fig.~(\ref{fig:twophotonstates}-a), to
the product state $\ket{\psi^{(2)}_p}$, Fig.~(\ref{fig:twophotonstates}-b), and to the spatially anti-correlated state 
$\ket{\psi^{(2)}_a}$, Fig.~(\ref{fig:twophotonstates}-c).
Note the relation between spatial and spectral distributions: the state 
$\ket{\psi^{(2)}_c}$, which is strongly correlated in space, Fig.~(\ref{fig:twophotonstates}-a), is
strongly anti-correlated spectrally Fig.~(\ref{fig:twophotonstates}-d), and vice versa for
$\ket{\psi^{(2)}_a}$. 
Irrespective of their correlation maps, all these states occupy the same spatial area on the upper-left edge 
of the lattice, see supplementary section I. The Schmidt number for $\ket{\psi^{(2)}_{c,a}}$ is $S_{N}=13$, while for 
$\ket{\psi^{(2)}_{p}}$ we have $S_{N}=1$. 
We now explore the robustness of our two-photon states as they traverse the disordered lattice. 
We begin with the product state $\ket{\psi^{(2)}_p}$. To characterize the impact of disorder, we 
compute the fidelity \cite{Nielsen2010} which is given as the overlap of the state $\ket{\psi^{(2)}_p(z_f)}$ after it has traversed the lattice 
with the reference state 
$\ket{\psi^{(2)}_p(z_m)}$ obtained after propagating the same state 
$\ket{\psi^{(2)}_p}$ in a disorder-free lattice.
The two wavepackets are taken at slightly different propagation distances $z_f$ and $z_m$ to account for the 
somewhat different travel distance in a disordered lattice. We find the fidelity
 $F_p=|\braket{\psi^{(2)}_p(z_f)|\psi^{(2)}_p(z_m)}|^2=0.98$, confirming that 
both the single-photon states and their product are immune to disorder. 
The edge-mode content of the evolved state is almost 100\%, 
$E_p=\sum_{n,m}^{\mathcal{E}\otimes \mathcal{E}}| \braket{\phi^{(2)}_{n,m}|\psi^{(2)}_p(z_f)} |^2=0.9934$.
The product state traverses the lattice without distortion, in spite of the degeneracy between the two-photon edge-edge and bulk-bulk states.

\begin{figure*}[h!]
\begin{center}
\includegraphics[width=0.8\linewidth]{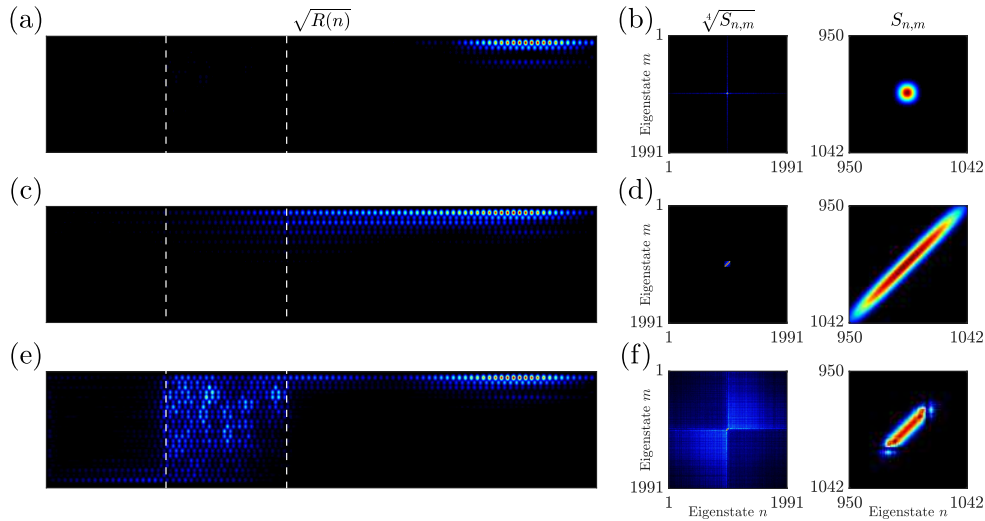}
\caption{Propagation of a two-photon edge state through a disordered topological lattice. Panel (a) shows the reduced single-photon
density distribution, and (b) the spectral correlation map for the product state $\ket{\psi^{(2)}_p}$,
which survives disorder. Note the rightmost panels in all cases show a magnification of the edge-edge subspace.
Panels (c, d) show the same for the spatially correlated entangled  state $\ket{\psi^{(2)}_c}$ in
the clean lattice, while panels (e, f) show the impact of disorder on this highly entangled
state. Dashed lines in (a, c, d) indicate the disordered region inside the lattice. 
}
\label{fig:anti}
\end{center}
\end{figure*}

Figs.~(\ref{fig:anti}-a,b) visualize this outcome by showing 
the single-photon spatial distribution and two-photon spectral correlation maps for the two-photon product state $\ket{\psi^{(2)}_p}$ traversing the disordered lattice. The single-photon 
spatial distribution $R(n)$ is given by the diagonal elements $\rho^{(1)}_{nn}$ of the 
reduced single-photon density matrix $\hat \rho^{(1)}$,
$R(n)\equiv \bra{n}\hat\rho^{(1)}\ket{n}\equiv \rho^{(1)}_{nn}$ \cite{Abouraddy2001}. The reduced single-photon density matrix 
$\hat \rho^{(1)}$ is obtained from the two-photon density matrix $\hat \rho^{(2)}$ in the usual
way, $\hat \rho^{(1)}=\sum_{m}^{M}\bra{m}\hat \rho^{(2)}\ket{m}$ \cite{Nielsen2010}. As expected, the spectral composition of the wavepacket
remains undisturbed and the wavepacket propagates through the disordered 
region without leaving the edge. 
We now turn our attention to entangled two-photon states. Figs.~(\ref{fig:anti}-c,d) depict 
$R(n)$ and the spectral correlation maps for the 
two-photon state $\ket{\psi^{(2)}_c}$ traversing the 
``clean'' lattice, and Figs.~(\ref{fig:anti}-e, f) show the same for the disordered lattice. While in the absence of
disorder, $R(n)$ stays on the edge and the highly correlated two-photon spectral distribution is 
unchanged (panels c, d), the disorder strongly affects these states. 
Fig.~(\ref{fig:anti}-e) shows strong dissipation into the bulk as soon
as the entangled wavepacket encounters 
the disordered region.
The spectral distribution spreads all over the system, with
both bulk-bulk and bulk-edge states becoming occupied, Fig.~(\ref{fig:anti}-f). A similar result is obtained for  
$\ket{\psi^{(2)}_a}$, except that the cross-like 
shape observed in Fig.~(\ref{fig:anti}-f) is flipped
towards the opposite diagonal, Supplementary section III.

To quantify the probability fraction of the states scattered into the bulk we compute the edge-mode content. For 
$\ket{\psi^{(2)}_c}$ the edge-mode content after traversing 
the disordered lattice is $E_c=0.4524$, while for $\ket{\psi^{(2)}_a}$ it gives $E_a=0.4453$.
Thus, more than 50\% of both types of states is scattered into the bulk.  The part of the
states that survives the disordered region and stays on the edge remains strongly correlated in the spectral domain: the edge-edge part of its spectral content 
preserves the initial shape, see the right column in Fig.~(\ref{fig:anti}-f).
However, the spectral phase of the state is scrambled. To illustrate this point, we have renormalized 
the transmitted edge part of the two-photon wavepacket to unity and computed its 
fidelity $F_N$ by overlapping it with the reference two-photon wavepacket from a clean system, yielding $F_N=0.405$.\\

\begin{figure}[t!]
\begin{center}
\includegraphics[width=0.7\linewidth]{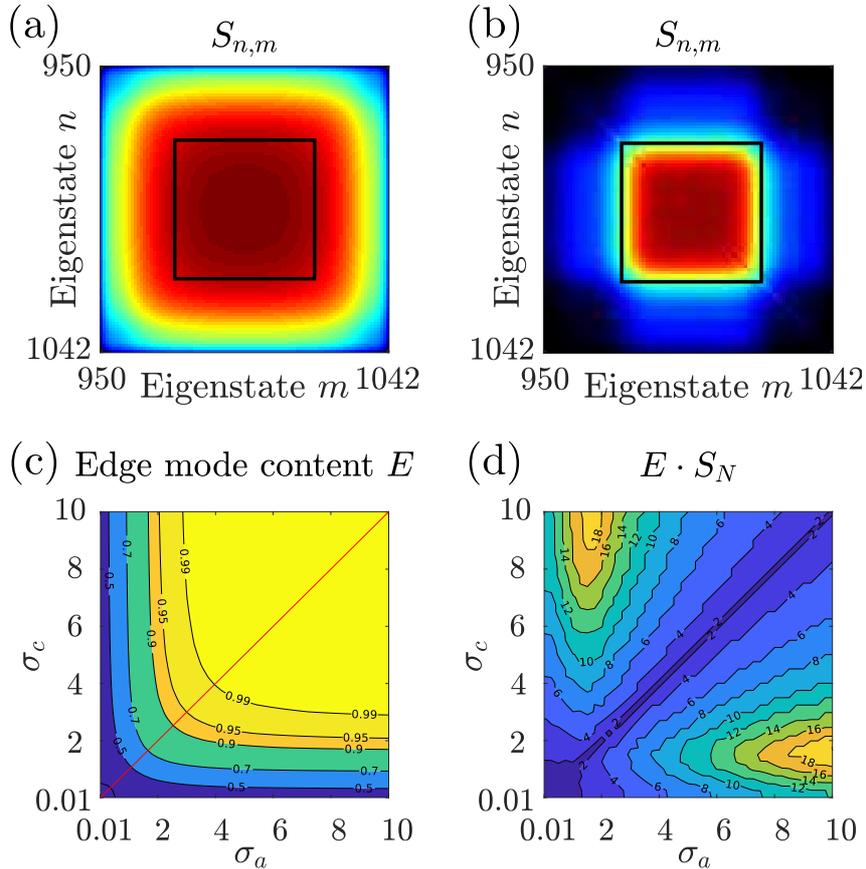}
\caption{
(a) In order to identify the topological protection window, we considered a spectrally broad product state $(\sigma_a=\sigma_c=0.01)$ as initial state for an ensemble of 1000 random Haldane lattices. In (b) we show the ensemble-average of the spectral correlation maps inside the edge-edge subspace after the propagation through the ensemble. We find that the only two-photon amplitudes that survive the disorder lie in the region indicated by the black square which is the protection window. In (c) we depict the edge-mode content, E. In (d) we have the product $E\cdot S_N$ as a function of $\sigma_c$ and $\sigma_a$ in the range  (0.01, 10).}
\label{fig:schmidt}
\end{center}
\end{figure}

We find that the conduit for dissipation of the two-photon edge-edge states 
is always provided by the edge-bulk states, which are degenerate in energy
with the edge-edge states. Once disorder induces 
transitions into the edge-bulk states, they further transfer the amplitudes 
into the energy-degenerate bulk-bulk states. 
Hence, the key to topological protection is to 
minimize the disorder-induced overlap of the single-photon edge states 
with the bulk, keeping the single-photon states
comprising the entangled state as close to the center of the gap as possible. 
That is, there is a topological protection window for single-photon states that offers the key guideline for designing robust two-photon states.
To infer the protection window, we sent a probe product state with $\sigma_c=\sigma_a=0.01$ through an ensemble of 1000 disordered lattices.
This initial state is very well localized to the edge region in real space, ensuring that
all components within the state travel along very close paths. 
The spectral content of the state before and after the disorder is shown in in Fig.~(\ref{fig:schmidt}-a, b). The components that have survived the impact of disorder are
within the marked window - the topological window of protection. Any entangled state with varying 
$\sigma_a$ and $\sigma_c$ must fit inside this protection window to be 
robust against disorder.

In practice, to increase the amount the entanglement we need to increase $\sigma_a$ $\left(\sigma_c\right)$ while decreasing $\sigma_c$ $\left(\sigma_a\right)$, and by doing so the wavefunction unavoidably tends to fall outside the protection window. However, we can always find two-photon states with a considerable amount of entanglement which are protected. To elucidate this we have scanned the edge mode content of the two-photon states after propagation through the disorder region as a function of $\sigma_a$ and $\sigma_c$. In Figs. (\ref{fig:schmidt} c-d) we show the contour maps of the edge-mode content as we vary $\sigma_a$ and $\sigma_c$.  Fig.~(\ref{fig:schmidt}-c)  shows the edge-mode content of the two-photon states
after propagation through the disordered region, with 
the diagonal corresponding to the product states, that is, states with $\sigma_a=\sigma_c$. 
The states with the highest degree of entanglement correspond to
very different $\sigma_a$ and $\sigma_c$ and therefore they are found 
in the top left and lower right corners in Fig. (\ref{fig:schmidt}-c).  
In general, highly entangled states lay on regions with $\sigma_a\ll \sigma_c$ (top left corner)
or  $\sigma_a\gg \sigma_c$ (bottom right corner) and  
the edge-mode content quickly drops below 0.5. The reason is because as one increases $\sigma_a$, or $\sigma_c$, 
the tails of the spectral correlation ellipse fall outside of the protection window
and, as a result, the states scatter into the bulk. Similarly, uncorrelated states may experience the same fate when they are initially confined into a small spatial region, which is the case for states with $\sigma_a=\sigma_c\in\left(0, 2.5\right)$.
Fig. (\ref{fig:schmidt}-d) shows the key figure of merit, $E \cdot S_{N}$, the product of the Schmidt
number $S_{N}$ and the edge-mode content $E$.
The bright yellow islands indicate the best two-photon states which
combine robustness against disorder with high degree of entanglement.
Importantly, the spectral correlation ellipse of these states 
always fits into the protection window shown in Fig. (\ref{fig:schmidt}-b). 
It is worth mentioning the features exhibited by the contour maps are generic as similar structures are obtained for disordered Haldane lattices with different dimensions, see Supplementary section IV. This demonstrates that, in principle, one can create states with high Schmidt number and edge-mode content close to unity.\\
As evidence that our results are generic, in the sense that they apply to other 2D topological systems, in the supplementary section V we have performed a similar analysis for an aperiodic topological lattice system \cite{Hafezi2013,HafeziNatPhys}. We have found that the contour map of the edge mode content $E$ is not symmetric, implying that the correlated states are slightly better protected than their anti-correlated ``mirror-images". Nevertheless, we obtain the same qualitative features as in the Haldane model.\\ 
%
Before concluding, we would like to outline possible ways to generate the initial states and address the potential challenges for experimental observations of these effects. The initially highly correlated states can be implemented using standard spontaneous-parametric-down-conversion nonlinear crystals to generate photon pairs that are coupled to the edge of the lattice using a positive achromatic doublet lens as demonstrated in \cite{Giuseppe2013}. Anticorrelated photon pairs can be generated by applying the fractional Fourier transform to the highly correlated states \cite{Weimann2016}. 
The Haldane lattice has been previously demonstrated using femtosecond laser written waveguides as reported in \cite{Rechtsman2013}. Hence, the challenges are reduced to optimizing the fabrication for minimal scattering, absorption and bending losses associated with the helical waveguides.

These results lead to the following conclusions. Two issues have to be considered
when constructing two-photon entangled edge states in topological systems: their
dissipation into the bulk and the relative dephasing between the different
components comprising the entangled state. 
Regarding dissipation, the two-photon edge states are protected just as well 
as the single-photon edge states. Further, phase scrambling can also be minimized if the different components of the entangled state travel along
the same path in the edge region. Both aims are 
achieved by keeping the spectral correlation map of 
the two-photon state in the center of the bandgap. 
Thus, attempts to increase entanglement
must be balanced against keeping the spectral correlation
maps of the two-photon states within the narrow spectral
region at the very center of the single-photon gap - the topological window of protection. This limits
the degree of entanglement one can safely encode in practice, but presents a 
clear strategy for creating useful states with high degree of
entanglement and robustness. 

Looking forward,  one could take advantage of 
the static nature of disorder to circumvent  
entanglement-induced dissipation into the bulk.  
While the disorder-induced relative phase between the different 
product-state components of the entangled wavepacket may
appear random due to the random nature of disorder,  for static disorder 
scrambling and dissipation are nevertheless fixed. 
This opens an opportunity to find the windows of protection
as we have done in the cases considered here, and generate robust wavepackets
tailored to the particular disordered system at hand. From a practical
perspective, the stability of entangled states up to relatively
high Schmidt numbers offers practical guidelines for generating
useful entangled edge states in topological photonic systems. Finally, our work may open the door to study topological protection of highly-entangled multiphoton non-Gaussian states  that fulfill the protection conditions.
\newpage
\section{Supplementary Information:}
\subsection{I Construction of the initial two-photon states}
\begin{figure}[b!]
\begin{center}
\includegraphics[width=0.85\linewidth]{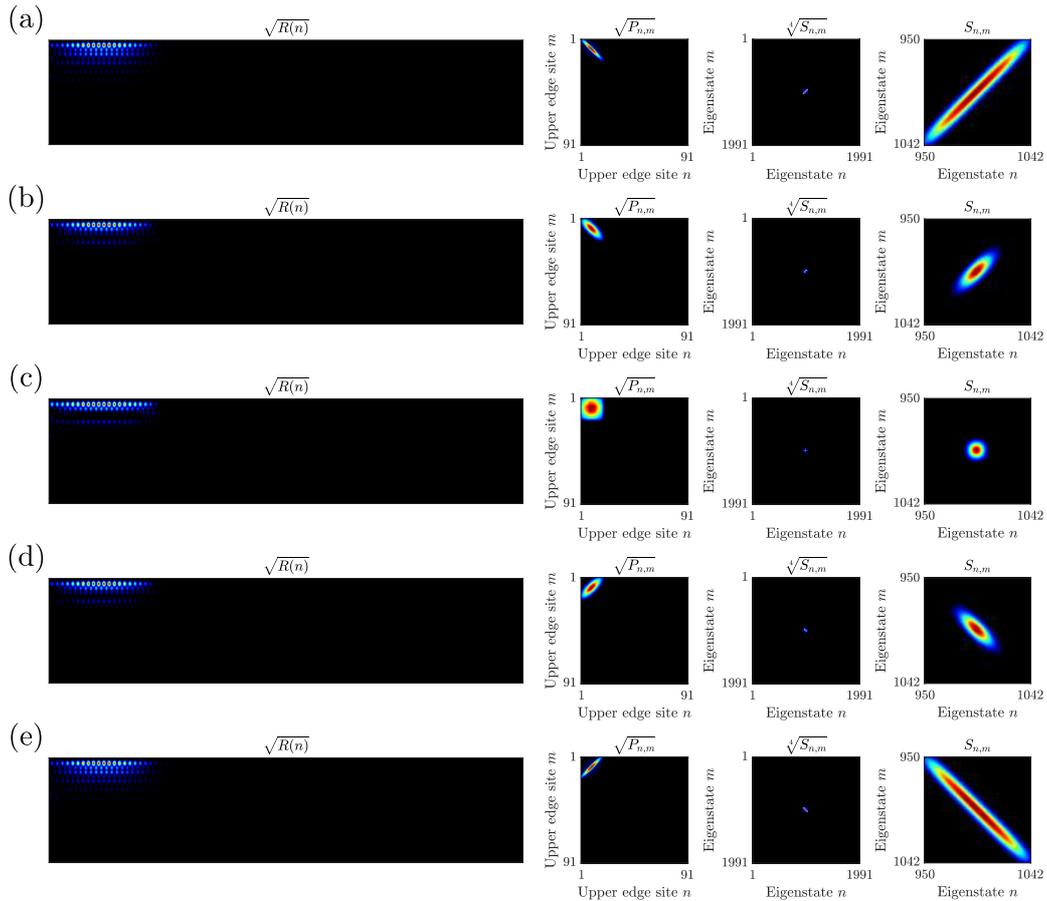}
\caption{Plot of the reduced density matrix representation $R(n)$, the spatial correlation map $P_{n,m}$, the spectral correlation map $S_{n,m}$ for the five different states considered in our simulations. (a) Correlated state, (b) semi-correlated state, (c) correlated state, (d) semi-anticorrelated state, and (e) anticorrelated state. Note that we show the square-root/fourth-root - as indicated above the panels - in order to increase the visibility of components with small probability.}
\label{fig:initialstates}
\end{center}
\end{figure}

As explained in the main text, we constructed the initial states by choosing different values for $\sigma_c$ and $\sigma_a$ in the expression
\begin{equation}
\ket{\tilde{\psi}^{(2)}_{\sigma_c,\sigma_a}}=\sum_{j,k=1}^{M_e} \psi_{j,k}\ket{j,k}=\sum_{j,k=1}^{M_e}(-1)^{j+k} e^{-\frac{(j-k)^2}{4 \sigma_a^2}-\frac{\left(x_0-(j+k)/2\right)^2}{\sigma_c^2}}\ket{j,k},
\label{eq:EG}
\end{equation}
which we then project onto the $\mathcal{E}\otimes\mathcal{E}$-subspace and renormalize the resulting state. As examples, we obtain the correlated state $\ket{\psi^{(2)}_c}$ ($\sigma_c=\sqrt{40}$, $\sigma_a=0.01$), the semi-correlated state $\ket{\psi^{(2)}_{sc}}$ ($\sigma_c=\sqrt{40}$, $\sigma_a=\sqrt{40}/3$), the product state $\ket{\psi^{(2)}_p}$ ($\sigma_c=\sqrt{40}$, $\sigma_a=\sqrt{40}$), the semi-anticorrelated state $\ket{\psi^{(2)}_{sa}}$ ($\sigma_c=\sqrt{40}/3$, $\sigma_a=\sqrt{40}$) and finally the anti-correlated state $\ket{\psi^{(2)}_{a}}$ ($\sigma_c=0.01$, $\sigma_a=\sqrt{40}$). In Fig.~(\ref{fig:initialstates}) we show the spatial and spectral correlation maps as well as the reduced density matrix representations of these states. As one can see, despite the fundamentally different correlation maps, all states occupy the same spatial region on the upper left edge of the lattice.
\newpage
\subsection{II Propagation in the clean system}

\begin{figure}[b!]
\begin{center}
\includegraphics[width=0.8\linewidth]{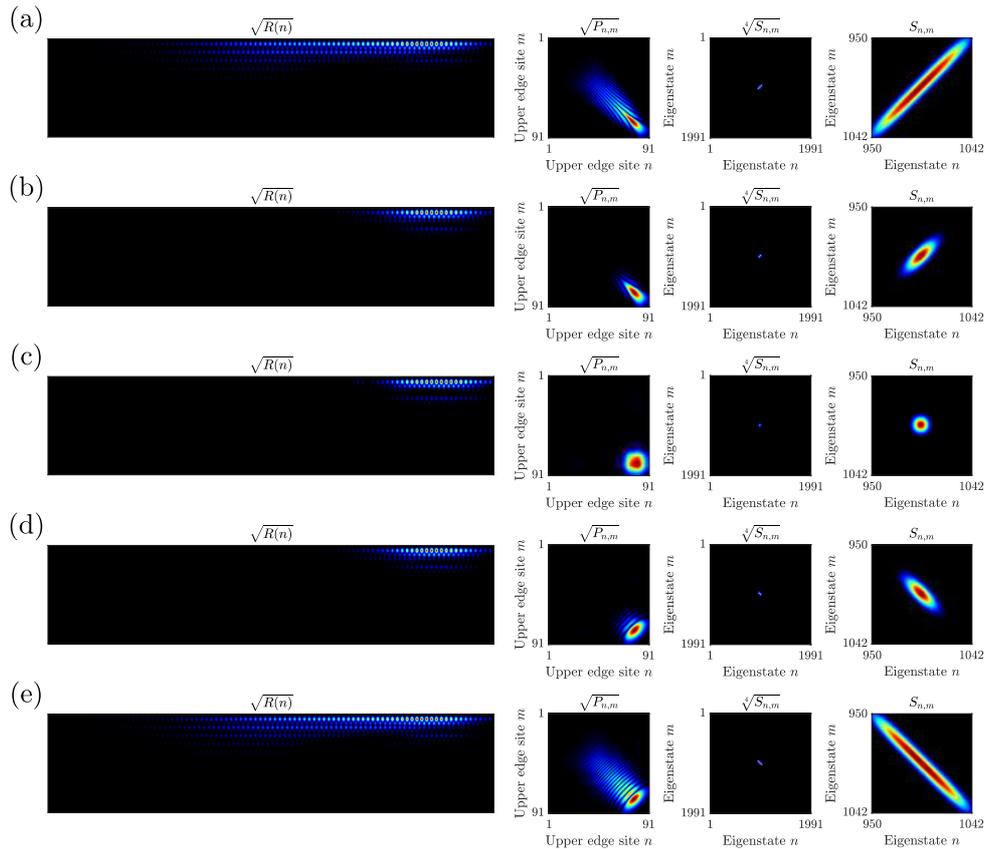}
\caption{Reduced density matrix $R(n)$, spatial $P_{n,m}$, and spectral $S_{n,m}$ correlation maps for the five states considered in our simulations after propagation distance $z_c=75$ in a clean lattice. (a) Correlated state (b) semi-correlated state (c) correlated state (d) semi-anticorrelated state (e) anticorrelated state. To increase the visibility of components with small probability we show the square-root/fourth-root.}
\label{fig:cleanprop}
\end{center}
\end{figure}

In order to assess the impact of disorder on two-photon states, we first analyze their propagation in a disorder-free lattice. Fig.~(\ref{fig:cleanprop}) depicts the final states after a propagation distance $z_c=75$. Even though there is no disorder present in the system, in the two cases, $\ket{\psi^{(2)}_c}$ and $\ket{\psi^{(2)}_a}$, the spatial probability distribution tends to spread out over the edge of the lattice losing the initial Gaussian shape, as illustrated in Figs.~(\ref{fig:cleanprop} a-e). In terms of correlations, both states spread out towards the four corners of the spatial correlation map. However, as the spreading is more prominent along the main diagonal we assert that in both cases the photons tend to bunch into the same site. Concurrently, we observe the emergence of interference fringes parallel (orthogonal) to the main diagonal of the correlation map corresponding to the correlated (anticorrelated) state. This implies that certain states are suppressed as a result of destructive quantum interference. Naturally, the spectral correlation maps remain invariant upon propagation, as well as the edge-mode content $E_p=1$ ensuring that only edge modes are present in the evolved wave-packet.\\
It is important to stress that the broadening of the wavefunctions is due to multimode interference, and since the two-photon eigenmodes exhibit larger propagation eigenvalues, the spreading rate is faster compared to single-photon wavepackets. In view of the spatial distortions undergone by $\ket{\psi^{(2)}_c}$ and $\ket{\psi^{(2)}_a}$, one can directly state that entangled states degrade even in disorder-free Haldane topological lattices. This intrinsic dispersion of strongly entangled states - even without disorder - may pose an additional challenge for their application in topological quantum information processing.\\
%
\subsection{III Propagation through disordered lattices}
\begin{figure}[b!]
\begin{center}
\includegraphics[width=0.8\linewidth]{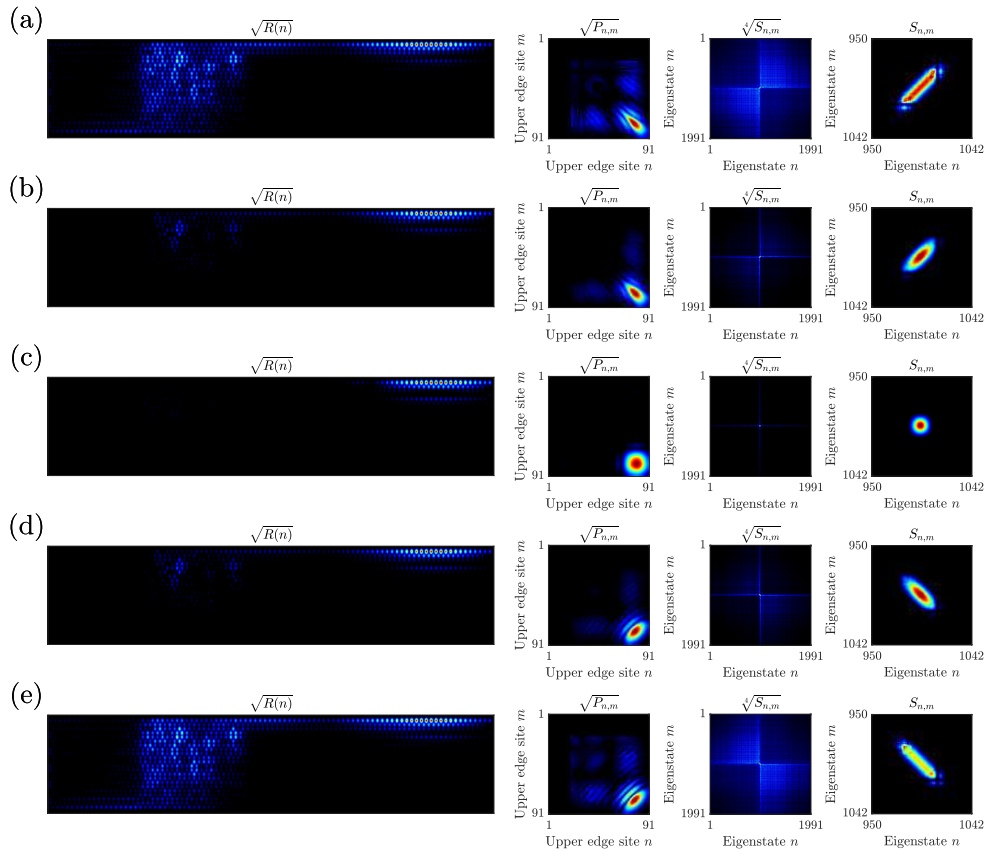}
\caption{Structure of the reduced density matrix $R(n)$, the spatial $P_{n,m}$, and the spectral $S_{n,m}$ correlation maps for the five different states considered in our simulations. The propagation distance $z_d=78.5$. In (a) we show the results for the correlated (b), semi-correlated (c), correlated (d), semi-anticorrelated, and (e) anticorrelated states. Note that we are showing the square-root/fourth-root - as indicated above the panels - in order to increase the visibility of components with small probability.}
\label{fig:disorderprop}
\end{center}
\end{figure}

Here we present the resulting states after the propagation through disorder. The correlated $\ket{\psi^{(2)}_c}$ and anticorrelated $\ket{\psi^{(2)}_a}$ two-photon states scatter significantly into the bulk of the disordered region. In the first place, spatial correlations, Figs.~(\ref{fig:disorderprop}-a) and (\ref{fig:disorderprop}-e), present some notable differences with their counterparts obtained in the clean system, Figs.~(\ref{fig:cleanprop}-a) and (\ref{fig:cleanprop}-e). That is, in the disordered cases the correlation maps no longer broaden along the main diagonal but they expand away of it redistributing the probabilities into three lobes. Yet, the highest probabilities are localized in the central lobe indicating that the photons remain  mainly correlated and anticorrelated as the corresponding initial states. Accordingly, in the spectrum the wavefunctions turn to be wider as demonstrated by the correlation maps shown  in the third columns of Figs.~(\ref{fig:disorderprop}-a) and (\ref{fig:disorderprop}e). A closer look into the central probability lobes, shown in the right-most column, reveals that the spectrally anticorrelated and correlated nature of the initial states survive the impact of disorder to some extent. Indeed, by monitoring the full dynamics one can see how the wavefunctions lose their correlation properties upon scattering and eventually the transmitted parts recover the initial correlation structure.\\ 
\clearpage 
\subsection{IV Effects of the lattice size}

\begin{figure}[h!]
\begin{center}
\includegraphics[width=\linewidth]{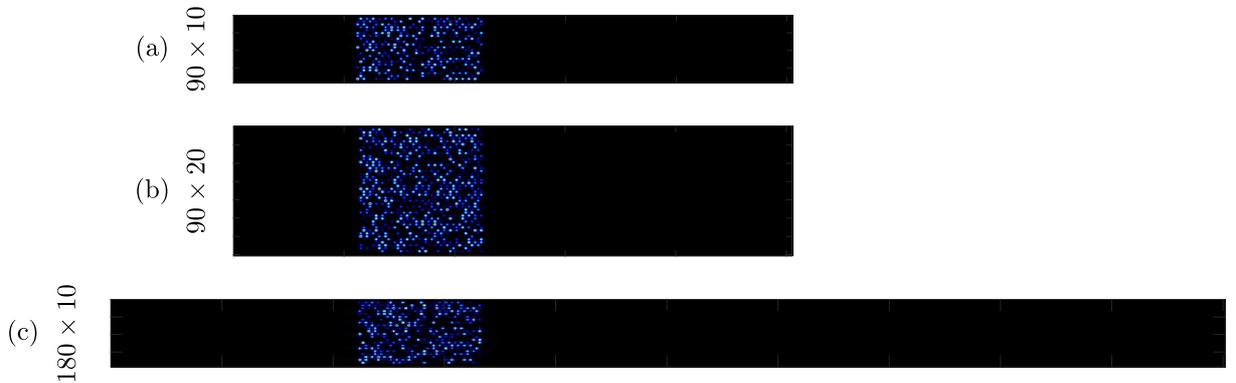}
\caption{Sketch of the lattices considered in the analysis of the impact of the system size. (a) Original lattice from the main text. (b) Lattice with twice the width. (c) Lattice with twice the length. In all cases, the disordered region has the same length $N_d=20$.}
\label{fig:lattices}
\end{center}
\end{figure}

\begin{figure}[h!]
\begin{center}
\includegraphics[width=.7\linewidth]{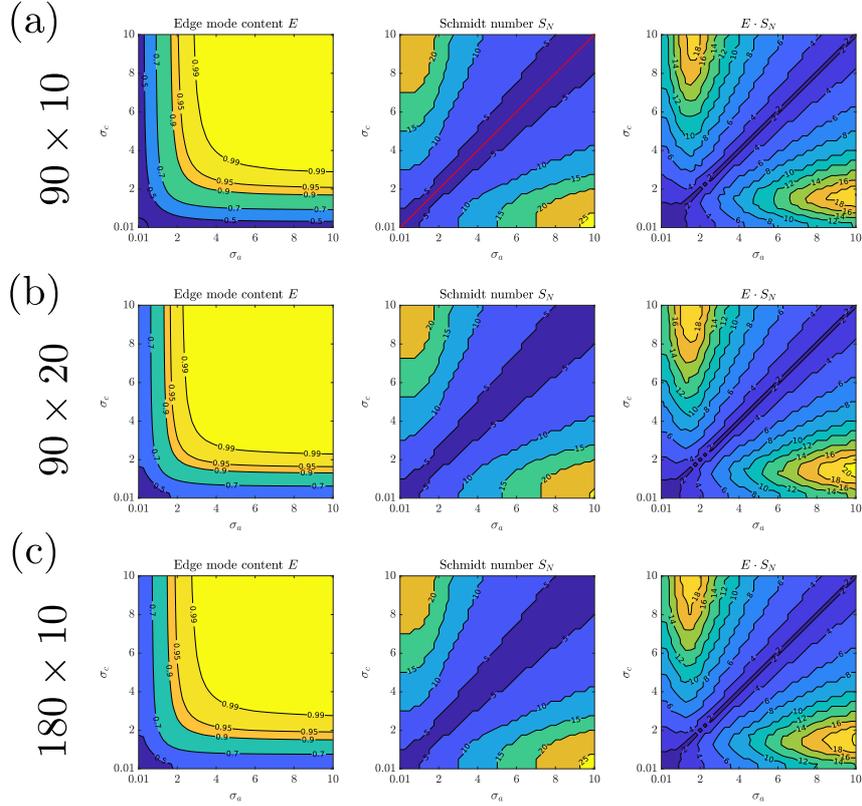}
\caption{Results of the parameter scans of $(\sigma_a,\sigma_c)$. The columns correspond to (from the left to the right) the edge-mode content $E$, Schmidt number $S_N$ and the combined figure of merit $E\cdot S_N$. (a) Original lattice. (b) Double width. (c) Double length. All contour maps display the same features, where highly entangled states (close to the $\sigma_a$-/$\sigma_c$-axis) are highly impacted by disorder.}
\label{fig:scans}
\end{center}
\end{figure}

\begin{figure}[h!]
\begin{center}
\includegraphics[width=.7\linewidth]{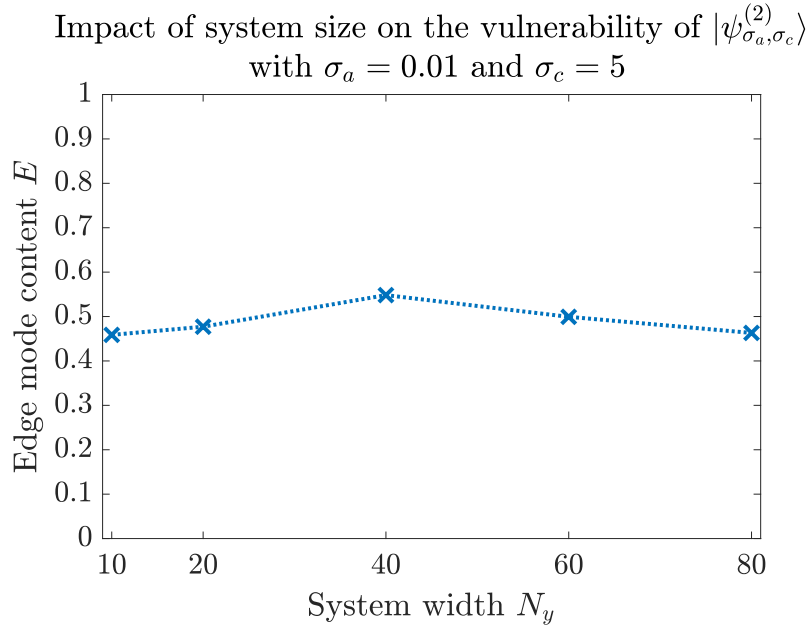}
\caption{Plot of the edge-mode content $E$ of the correlated state $\ket{\psi^{(2)}_{\sigma_a,\sigma_c}}$ after the disorder, with $\sigma_a=0.01$, $\sigma_c=5$, against increasing widths of the lattice $N_x=90, N_y=10,20,40,60,80$. As one can see, $E$ remains close to 0.5 in all cases, as a result, one can conclude that the impact of disorder on highly entangled states is independent of the lattice size .}
\label{fig:largersystems}
\end{center}
\end{figure}

We now explore how the size of the Haldane lattice influences the protection window. To do so, we consider two additional lattices with double spatial length ($N_x=10, N_y=180$ hexagons) and width ($N_x=20,N_y=90$). In both cases the length of the disordered region ($N_d=20$ hexagons in $y$-direction) is the same as in the original lattice ($N_x=10,N_y=90$), see Fig.~(\ref{fig:lattices}). The parameter scans in Fig.~(\ref{fig:scans}) yield, essentially, the same contour maps for the edge-mode content $E$ and the product $E\cdot S_N$. To further corroborate this finding, we tested even larger systems ($N_x=90,N_y=10,20,40,60,80$), for a correlated state with $\sigma_a=0.01$ and $\sigma_c=5$. As depicted in Fig.~(\ref{fig:largersystems}), the edge-mode content of this state, after the disordered region, is also independent of the system size.  Accordingly, we conclude that the results discussed in the main text are generic and not a mere effect of the system size. 

\clearpage
\subsection{V Topological protection of entangled two-photon states in an aperiodic topological insulator}

\begin{figure}[h!]
\begin{center}
\includegraphics[width=.8\linewidth]{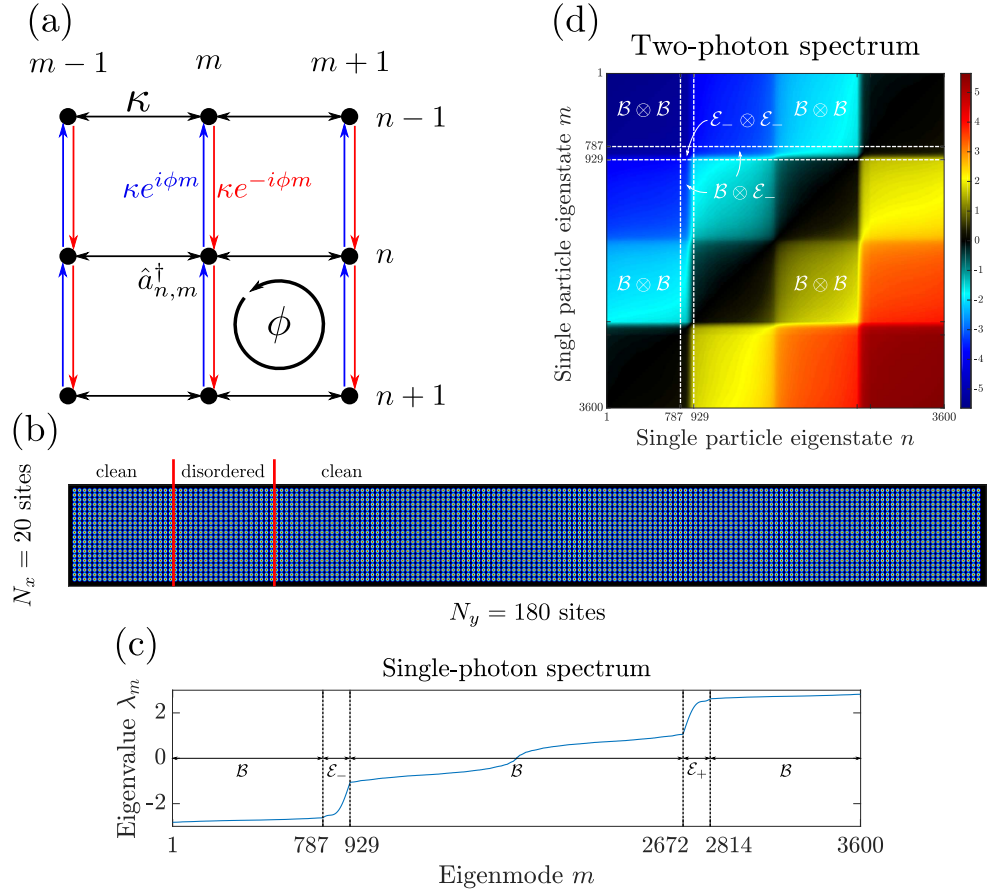}
\caption{\textbf{(a)} Sketch of the coupling structure in the aperiodic topological insulator. \textbf{(b)} Finite ribbon considered in our simulations. \textbf{(c)} Single-photon spectrum. \textbf{(d)} Two-photon spectrum.}
\label{fig:qhe}
\end{center}
\end{figure}

In order to show that our results are applicable to other 2D topological systems - and to connect them with established experimental work in photonic systems, we now consider the evolution of two-photon states in a topological insulator based on an aperiodic lattice. Such a system has been implemented experimentally using aperiodic networks of coupled ring-resonators \cite{Hafezi2013,Hararieaar4003}. At the single-particle level, this system is described by the Hamiltonian
\begin{equation}
\hat{H}=\kappa \sum_{n,m} \hat{a}^\dagger_{n,m}\hat{a}_{n,m+1} + e^{-i\phi m} \hat{a}^\dagger_{n,m} \hat{a}_{n+1,m} + h.c.
\end{equation}
The site modes represented by the operators $\hat{a}^\dagger_{n,m}$ form a 2D-square lattice with nearest-neighbor hopping, where the coupling $\kappa$ in $y$-direction (left to right edge, index $m$) is real-valued. In the $x$-direction (top to bottom edge, index $n$) the sites exhibit a complex-valued coupling $\kappa e^{-i\phi m}$, which is dependent on the $y$-coordinate $m$, as sketched in Fig.~(\ref{fig:qhe}-a). Specifically we choose $\phi=\frac{\pi}{2}$, which ensures that the phase accumulated around any local 4-site plaquette is $-\phi m + \phi (m+1) = \phi= \frac{\pi}{2}$. For our simulations we consider a finite ribbon with $N_x \times N_y = 20 \times 180$ sites, Fig.~(\ref{fig:qhe}-b), where the vertical dashed lines indicate the region where we introduce static disorder. The single-photon spectrum (without disorder) features two disjoint edge-spaces $\mathcal{E}_{\pm}$, which correspond to clock-wise (CW, $\mathcal{E}_+$) and counter-clockwise (CCW, $\mathcal{E}_-$) propagating edge-modes, Fig.~(\ref{fig:qhe}-c). As we have done for the Haldane lattice, we prepare states that start on the top-left edge of the system. As such, it is convenient to project them only onto the $\mathcal{E}_-$-subspace, where the states then propagate CCW directly into the disordered region. Thus we define the $\mathcal{E}_+$-space to be part of the bulk space $\mathcal{B}$, which ensures that - despite dissipation effects - also back-scattering is reflected in the edge-mode content of the states after propagation through the disorder. We show the resulting two-photon spectrum (without disorder) in Fig.~(\ref{fig:qhe}-d), which indicates again the massive degeneracies between the $\mathcal{B}\otimes \mathcal{B}$, $\mathcal{B}\otimes\mathcal{E}$ and $\mathcal{E} \otimes \mathcal{E}$ two-photon subspaces.\par
\begin{figure}[t!]
\begin{center}
\includegraphics[width=.7\linewidth]{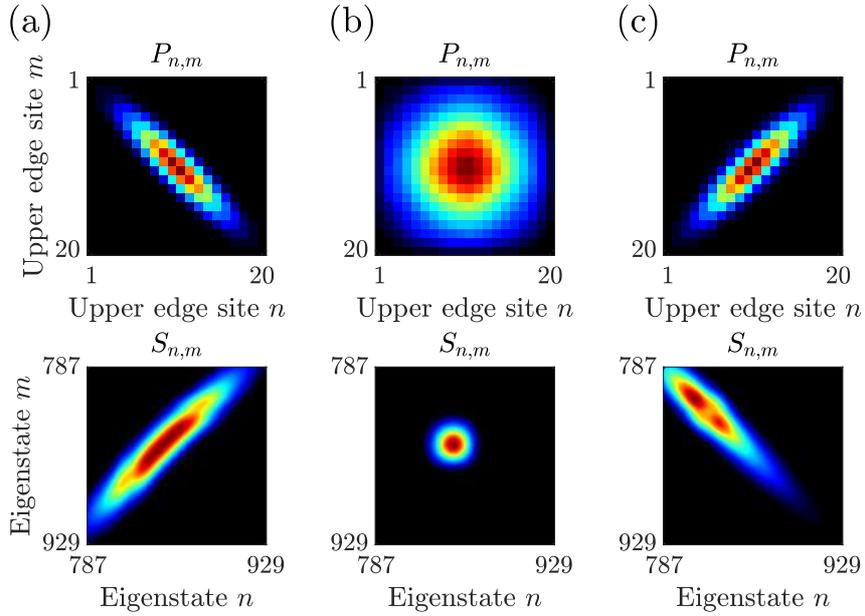}
\caption{\textbf{(a)} Spatially correlated state with $\sigma_a=1.2, \sigma_c=6$. \textbf{(b)} Product state with $\sigma_a=\sigma_c=6$. \textbf{(c)} Spatially anti-correlated state with $\sigma_a=6, \sigma_c=1.2$. \textbf{Top:} Spatial correlation map $P_{n,m}$ of the first 20 sites on the top-left edge of the system. \textbf{Bottom:} Spectral correlation map $S_{n,m}$ in the edge-edge subspace.	}
\label{fig:qhe-states}
\end{center}
\end{figure}

We construct the two-photon states in the same way as for the Haldane lattice using the template states
\begin{equation}
\ket{\psi^{(2)}_{\sigma_c,\sigma_a}} = \sum_{j,k=1}^{M_e} (-i)^{j+k} e^{-\frac{(j-k)^2}{4 \sigma_a^2}-\frac{(x_0-(j+k))^2}{\sigma_c^2}} \ket{j,k},
\end{equation}
but now we require the local phases $(-i)^{j+k}$ to obtain edge-states with the proper CCW-momentum. After projection onto the $\mathcal{E}\otimes \mathcal{E}$-subspace, and renormalization, we obtain the exemplary states shown in Fig.~(\ref{fig:qhe-states}). In comparison to the Haldane lattice, these states display very similar spatial correlations (by construction) but significantly different spectral correlation maps. Specifically, their spectral correlation ellipses are not oriented around the center of the edge-edge subspace. This is a consequence of the aperiodic nature of the lattice, which induces an asymmetric dispersion relation with respect to the center of the $\mathcal{E}_-$ ($\mathcal{E}_+$) subspace in the single-photon spectrum, Fig.~(\ref{fig:qhe}-c). However, we still observe that the spatially correlated state is spectrally anti-correlated, and vice versa for the spatially anti-correlated state. \par
\begin{figure}[h!]
\begin{center}
\includegraphics[width=.7\linewidth]{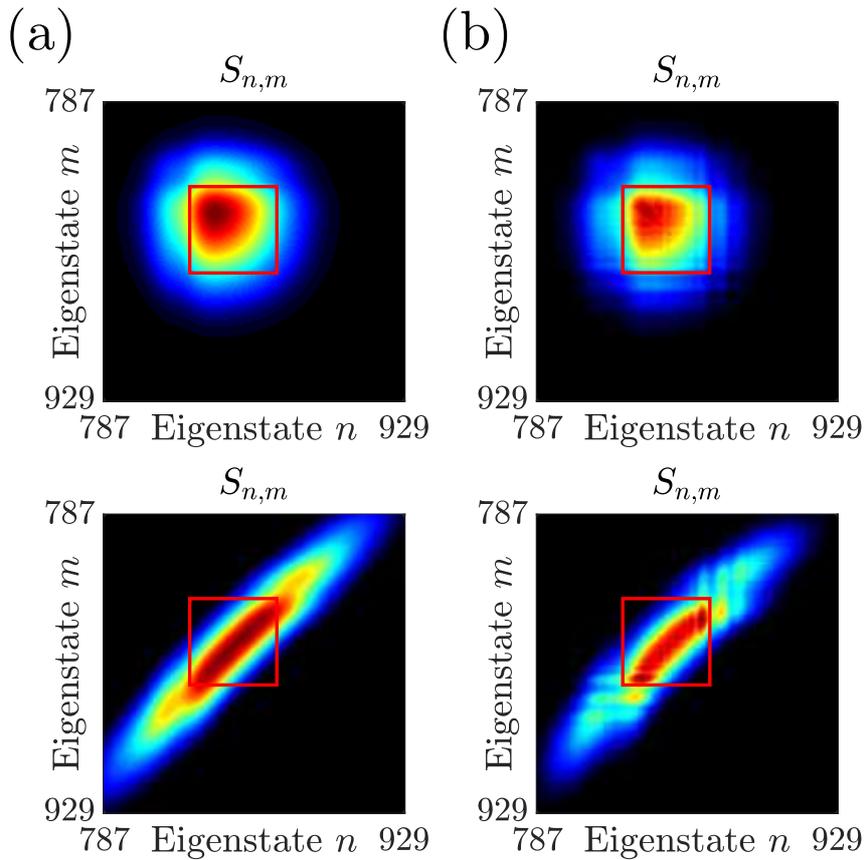}
\caption{\textbf{(a)} Initial states to probe the window of protection. \textbf{(b)} Ensemble average spectral correlation map after propagation through 200 instances of the random disorder. \textbf{Top:} Product state with $\sigma_a=2.5=\sigma_c$. \textbf{Bottom:} Correlated state with $\sigma_a=1.2$, $\sigma_c=6$.}
\label{fig:qhe-window}
\end{center}
\end{figure}
\begin{figure}[t!]
\begin{center}
\includegraphics[width=\linewidth]{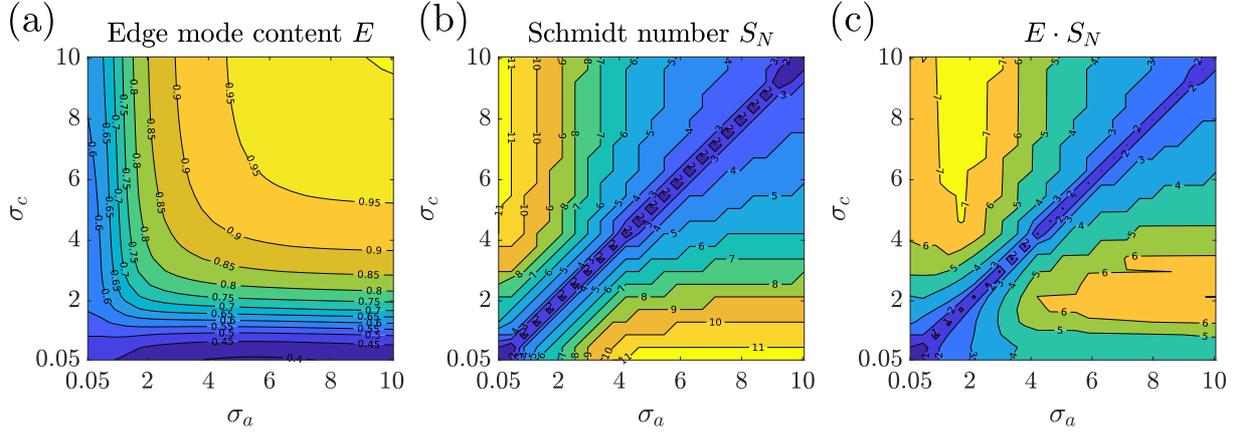}
\caption{\textbf{(a)} Contour-plot of the edge-mode content $E$ of the states $\ket{\psi^{(2)}_{\sigma_c,\sigma_a}}$ after propagation through the disorder (disorder strength $\sigma=0.3$, propagation distance is $z=450$). \textbf{(b)} Schmidt-numbers $S_N$ of the initial two-photon states $\ket{\psi^{(2)}_{\sigma_c,\sigma_a}}$. \textbf{(c)} The figure of merit $E \cdot S_N$.}
\label{fig:qhe-scan}
\end{center}
\end{figure}

In order to identify the window of protection, we launch the spectrally wide product-state $\ket{\psi^{(2)}_{\sigma_c=2.5,\sigma_a=2.5}}$ and the correlated state $\ket{\psi^{(2)}_{\sigma_c=6,\sigma_a=1.2}}$ through an ensemble of $200$ instances of disordered lattices (strength of the disorder $\sigma=0.3$) and observe the surviving spectral amplitudes in Fig.~(\ref{fig:qhe-window}). Notably, also the window of protection does not lie in the center of the edge-edge subspace. However, we can deduce that more highly entangled states will be less protected in this system. This is indeed the case, as one can see in the parameter-scans over $(\sigma_c,\sigma_a)$ in Fig.~(\ref{fig:qhe-scan}). In complete analogy to the Haldane lattice, the most strongly entangled states - largest Schmidt-number $S_N$ - are close to the $\sigma_a$-/$\sigma_c$-axes, where also the edge-mode content $E$ after propagation through the disorder is low. Quite interestingly, we observe an asymmetry between correlated- and anti-correlated states, such that anti-correlated states ($\sigma_a>\sigma_c$) are protected to a lesser degree than their correlated ``mirror images'' ($\sigma_a<\sigma_c$). This is also a consequence of the aperiodicity of the lattice. From a spatial perspective, the photons in an anti-correlated state tend to occupy opposite sides of the wavepacket, whereby they experience different local coupling coefficients. On the other hand - in the correlated states - the photons tend to occupy the same site and experience the same coupling. From a spectral perspective, this can also be seen in Fig.~(\ref{fig:qhe-states}). If one starts with a product state $\sigma_a = \sigma_c$ - Fig.~(\ref{fig:qhe-states}-b) - and decreases $\sigma_c$ then the ``center of mass'' of the spectral correlation ellipse tends to shift towards the upper-left corner of the edge-edge subspace as, as seen in Fig.~(\ref{fig:qhe-states}-c). At the same time, the correlation ellipse widens along the diagonal but this is not compensated by the previously mentioned effect. Eventually the state occupies mostly eigenstates outside of the window of protection. On the other hand, when decreasing $\sigma_a$ - Fig.~(\ref{fig:qhe-states}-a) the center of mass of the spectral correlation ellipse tends to stay in place while it widens along the anti-diagonal. The net-effect is that spatially anti-correlated states are further away from the window of protection and thus more vulnerable than correlated states. \par
We stress that the dispersion of the two-photon wavepackets is significantly stronger in comparison to the Haldane lattice. Even though we choose a much longer lattice than in the Haldane case, the wavefunctions spread around the complete edge of the system even before the slower parts of the wavefunction can leave the disordered region. This poses a challenge to the comparability of the results. But nevertheless, we observe very similar features and conclude that our results also apply in the present aperiodic topological system.
\clearpage

\bibliography{literature}

\end{document}